\documentstyle[prl,aps,epsf,floats]{revtex}

\begin{document}

\twocolumn[\hsize\textwidth\columnwidth\hsize\csname @twocolumnfalse\endcsname
\author{A.M. Bratkovsky$^{1}$ and A.P. Levanyuk$^{1,2}$}
\address{$^{1}$ Hewlett-Packard Laboratories, 1501 Page Mill Road, Palo
Alto, California 94304\\
$^{2}$Departamento de F\'{i}sica de la Materia Condensada, C-III,
Universidad Aut\'{o}noma de Madrid, 28049 Madrid, Spain
}\title{
Smearing of phase transition 
due to surface effect  
in ferroelectric nanostructures
}
\date{January 30, 2004}
\maketitle
\begin{abstract}
The boundary conditions, customarily used in the 
Landau-type approach to ferroelectric thin films and nanostructures,
have to be modified to take into account that 
a surface of a ferroelectric (FE) is a defect of a ``field'' type.
The surface (interface) field  is coupled to a normal component of polarization
and,  as a result, the second order phase transitions 
are generally suppressed and anomalies in response 
are washed out, as observed experimentally.
\end{abstract}
\vskip2pc] \narrowtext
Theoretical studies of phase transitions in thin films and the corresponding
size effects within the Landau theory \cite{Landau37} have been undertaken
since 1950s. Recently, the interest to these questions has risen
dramatically in view of the applications of ferroelectric thin films \cite
{scott00} and a discovery of various ferroelectric nanostructures\cite
{nanofe}. The boundary conditions for thin films were originally discussed
by Ginzburg and Landau (GL) in 1950\cite{GinzburgLandau} and later by
Ginzburg and Pitaevskii \cite{GPit58}. It was shown by GL that, if the
properties of the boundary layer are the same as of the bulk, one arrives at
the condition that the gradient of the order parameter $\eta $ vanishes at
the surface, $\vec{\nabla}_{\vec{n}}\eta =0$ (in zero magnetic field, $\vec{n%
}$ is the normal to the surface). Starting from a microscopic theory, de
Gennes has shown that for a superconductor-metal interface with no current
and magnetic field a more general boundary condition applies, $\vec{\nabla}_{%
\vec{n}}\eta +\eta /\delta =0,$ where $\delta $ is the characteristic length
scale describing the proximity effect \cite{degennes}. These conditions are
very general and were introduced phenomenologically by Kaganov and
Omelyanchouk for a surface of a ferromagnet \cite{Kaganov} (cf. review in 
\cite{Binder72}). They assumed that the surface energy was $\propto \eta ^{2}
$. Kretschmer and Binder \cite{Binder79}, using the same boundary
conditions, have taken into account the depolarizing field, which is
important for a polarization (magnetization) perpendicular to a surface.
These boundary conditions are customarily used in studies of phase
transitions in ferroelectric films (see, e.g. \cite{Tilley2001}).

The choice of a surface energy dependence on the order parameter{\bf \ }must
be governed by symmetry considerations, as it is for a bulk energy. Surface
symmetry is different from a symmetry of the bulk, and this is of special
importance for ferroelectrics. A surface eliminates all symmetry elements
that change the direction of the normal to the surface. If the order
parameter (or one of its components) is a vector perpendicular to the
surface, the surface energy contains a linear ``field''\ term reminiscent of
a ``local field defect'' \cite{LevSig88}. The ``local field'' is absent for
superconducting or magnetic order parameters because surface does not break
either gauge or magnetic symmetry. A Coulomb dipole field (double layer),
contributing to the work function \cite{bardeen36,Monch} and\ structural
surface relaxation (see, e.g. \cite{vbilt01}) are different aspects of this
phenomenological ``field''. If the order parameter is not a vector, the
``field'' term may appear in special cases if the surface eliminates
appropriate symmetry elements. If the field term is absent, some higher
order terms, forbidden in the bulk, can be still allowed by the surface
symmetry. We illustrate the effect of new boundary conditions for thin film
of uniaxial FE. This is relevant also to perovskite ferroelectrics that are
cubic in the bulk but become tetragonal in thin films because of elastic
misfit with a substrate. Let the surface be perpendicular to the polar axis $%
z$. The Landau-Ginzburg-Devonshire free energy for the bulk between the
electrodes $1$ and 2 at $z=\mp l/2$ is\cite{Chensky82,BL} 
\begin{eqnarray}
F_{LGD} &=&\int_{2}^{1}dV\Bigl[\frac{1}{2}AP^{2}+\frac{1}{4}BP^{4}+\frac{1}{6%
}CP^{6}  \nonumber \\
&&+\frac{1}{2}g\left( \frac{dP}{dz}\right) ^{2}+\frac{1}{2}D\left( \nabla
_{\perp }P\right) ^{2}\Bigr],  \label{eq:Ft}
\end{eqnarray}
where $P\equiv P_{z},$ and one adds to it the surface energy which,
generalizing Ref.\cite{Kaganov}, that can be written in the form 
\begin{equation}
F_{s}=\int_{1+2}dS\left( \frac{1}{2}\alpha P^{2}-wP\right) ,  \label{eq:Fws}
\end{equation}
where $\alpha $ corresponds to a ``temperature''-like component of the
surface energy and $w$ to its ``field'' component. We obtain from Eqs.~(\ref
{eq:Ft})\ and (\ref{eq:Fws})\ the {\em new} {\em correct boundary conditions 
}for ferroelectrics in the same way as in Ref. \cite{Kaganov}, 
\begin{equation}
\alpha _{1(2)}P+(-)g\frac{dP}{dz}=w_{1(2)},\quad z=+(-)l/2.  \label{eq:BCsrf}
\end{equation}
In CGSE\ units, one can estimate that $\alpha \sim d_{at},$ where $d_{at}$
is the characteristic ``atomic'' length scale, on the order of the lattice
constant. The electric field at the surface \cite{bardeen36,Monch} is on the
order of $\Phi _{s}/d_{at}\sim 1$V/\AA\ $\approx 10^{8}$V/cm, where $q\Phi
_{s}\sim 4$eV\ is the typical workfunction for ferroelectrics \cite{scott00}%
. The surface field corresponds to a polarization, which is on the order of
an ``atomic'' polarization $P_{at}=q/d_{at}^{2}\sim 200\mu C/$cm$^{2}$, so
that $w\sim P_{at}d_{at}$. We expect that the structural relaxation
contribution to $w$ is of the same order of magnitude: from first-principles
calculations at the ideal surfaces of BaTiO$_{3}$ and PbTiO$_{3}$ the
surface polarization $P_{surf}\sim 10^{-1}P_{at}$ \cite{vbilt01}. Usually
the top and bottom electrode interfaces to the FE\ film are dissimilar and
thickness of the surface polarized layers is $\sim d_{at}.$ This is enough
to produce a considerable field in the film. Indeed, the field in the
surface double layer is $E_{1}=-4\pi P_{surf}\sim -4\pi P_{at}.$ If, for
example, we consider a short-circuited film with strongly asymmetric double
layers at the interfaces (so that the voltage drop on the second layer is
much smaller than the one on the first layer)\ the surface voltage drop $%
\sim E_{1}d_{at}$ must be compensated by the field in the bulk, $%
E_{bulk}=-E_{1}d_{at}/l=4\pi P_{at}/l.$ For example, in a film with a
thickness $l\sim 1000$\AA\ we find $E_{bulk}\sim P_{at}d_{at}/l\sim 300$%
kV/cm, which is a very strong field indeed. In a ferroelectric film this
field smears second order transitions and may smear a first order phase
transition. In a free standing FE\ film the role of electrodes would be
played by a surface conductivity and/or atmospheric ions. Note that is a
rare case of {\em symmetric} surfaces the surface dipoles produce no
electric field in the film and {\em no smearing} of the phase transition
occurs. The polarization $P\left( z\right) $ is found from the equation of
state for (\ref{eq:Ft}) and the Poisson equation. Assuming that there is no
external charge, we have 
\begin{eqnarray}
AP+BP^{3}+CP^{5}-gd^{2}P/dz^{2} &=&E,  \label{eq:Eqstate} \\
d\left( E+4\pi P\right) /dz &=&0,  \label{eq:Gauss} \\
(1/l)\int_{1}^{2}Edz &=&\left( \varphi _{1}-\varphi _{2}\right) /l\equiv
E_{0},  \label{eq:biasv}
\end{eqnarray}
where $E_{0}$ is the external electric field. We obtain from Eqs.~(\ref
{eq:Gauss}) and (\ref{eq:biasv}) 
\begin{equation}
E=E_{0}-4\pi \left[ P\left( z\right) -\bar{P}\right] ,  \label{eq:field}
\end{equation}
where the overbar means an average over the film, i.e. $\ \bar{f}%
=(1/l)\int_{1}^{2}dzf(z).$ One probes experimentally not the actual surface
polarization but the net polarization in the sample. Indeed, the
experimentally measures the surface charge $\sigma $, which is equal to the
normal component of the displacement field, $\sigma =D_{z}/4\pi =E_{0}/4\pi +%
\bar{P}.$ Therefore, we need to find an equation for the net polarization $%
\bar{P},$ which, from the symmetry arguments alone, should have the form 
\begin{equation}
\check{A}\bar{P}+b\bar{P}^{2}+\check{B}\bar{P}^{3}+f\bar{P}^{4}+\check{C}%
\bar{P}^{5}=E_{0}+W/l,  \label{eq:PbarGen}
\end{equation}
since the terms of all powers in $\bar{P}$ are allowed in a system with
asymmetric boundaries. Below, we will solve the system (\ref{eq:Eqstate})-(%
\ref{eq:field}) with the boundary conditions (\ref{eq:BCsrf}) and find the
coefficients in the generic equation of state (\ref{eq:PbarGen}). To find
the net polarization $\bar{P},$ we integrate the equation of state (\ref
{eq:Eqstate}) over the film with the use of the boundary conditions (\ref
{eq:BCsrf}): 
\begin{eqnarray}
&&A\bar{P}+B\bar{P}^{3}+C\bar{P}^{5}+(3B\bar{P}+10C\bar{P}^{3})\overline{%
\delta P^{2}}  \nonumber \\
&&+(B+10C\bar{P}^{2})\overline{\delta P^{3}}+5C\bar{P}\overline{\delta P^{4}}%
+C\overline{\delta P^{5}}  \nonumber \\
&=&E_{0}+R/l,  \label{eq:Pbar} \\
R &\equiv &w_{1}+w_{2}-\alpha _{1}P_{1}-\alpha _{2}P_{2},  \label{eq:R}
\end{eqnarray}
where $\delta P=P(z)-\bar{P}.$ To find $\delta P$ we subtract Eq.(\ref
{eq:Pbar})\ from Eq.(\ref{eq:Eqstate}) and obtain 
\begin{equation}
\tilde{A}\delta P+K_{2}+K_{3}+K_{4}+K_{5}-gd^{2}\delta P/dz^{2}=-R/l,
\label{eq:EqSLin}
\end{equation}
where the cumulants $K_{2}=(3B\bar{P}+10C\bar{P}^{3})(\delta P^{2}-\overline{%
\delta P^{2}}),$ $K_{3}=(B+10C\bar{P}^{2})\left( \delta P^{3}-\overline{%
\delta P^{3}}\right) ,$ $K_{4}=5C\bar{P}\left( \delta P^{4}-\overline{\delta
P^{4}}\right) ,$ $K_{5}=C\left( \delta P^{5}-\overline{\delta P^{5}}\right) $
can be neglected \ under conditions used below, $\tilde{A}=4\pi +A+3B\bar{P}%
^{2}+5C\bar{P}^{4}\approx 4\pi ,$ since $A=(T-T_{c})/T_{0}$ and for
displacive systems $T_{0}\sim 10^{5}$K, so we are in a regime where $\left|
A\right| \lesssim T_{c}/T_{0}\ll 1.$ The solution of the equation (\ref
{eq:EqSLin})\ is 
\begin{eqnarray}
\delta P &=&-R/(\tilde{A}l)+c_{1}\exp \left[ -\lambda (z+l/2)\right] 
\nonumber \\
&&+c_{2}\exp [-\lambda (l/2-z)],  \label{eq:dP}
\end{eqnarray}
where $\lambda =\left( \tilde{A}/g\right) ^{1/2}\sim d_{at\text{ }}^{-1}.$

The fact of appearance of the atomic length $1/\lambda \sim d_{at}$ in our
phenomenological treatment is important and requires an explanation.
Actually, it has been already found in Ref.\cite{Binder79}. Physically, it
appears because the spatial changes of polarization that we consider lead to
bound charges and strong electric (``depolarizing'') fields at the
interfaces. As a result, the characteristic length of the spatial change of
the polarization there is comparable not{\bf \ }to the correlation length
(the width of the domain wall) but to a much smaller length which happens to
be of an atomic order for typical values of the coefficient $g.$ Strictly
speaking, the present phenomenological derivation is valid for models with $%
\sqrt{g}>d_{at}.$ Real materials are at the boundary of applicability of the
theory (length scale $\sim d_{at}$). From the data by Stemmer et al. for 90$%
^{0}$ domain walls in PbTiO$_{3}$ \cite{StemmerPTO95} we can estimate $%
1/\lambda =(g/4\pi )^{1/2}=0.4-0.5$\AA . This suggests that PbTiO$_{3}$ is
not exactly the displacive type system, where our estimate applies, $%
1/\lambda \sim d_{at}\sim 1-3$\AA , but it is comparable. Therefore, the
results below are semiquantitative, with the exact values of the
renormalized coefficients in Eq.(\ref{eq:PbarGen}) to be found from a
microscopic theory.

We are interested in a solution accurate to (leading) terms linear in $1/l.$
We find in this approximation 
\begin{equation}
c_{1(2)}=\left( w_{1(2)}-\alpha _{1(2)}\bar{P}\right) /\left( \alpha
_{1(2)}+\lambda g\right) ,  \label{eq:C12}
\end{equation}
and obtain from the condition $\int \delta Pdz=0$ 
\begin{equation}
R=\lambda g\left( \frac{w_{1}-\alpha _{1}\bar{P}}{\alpha _{1}+\lambda g}+%
\frac{w_{2}-\alpha _{2}\bar{P}}{\alpha _{2}+\lambda g}\right) .
\end{equation}
The averages in Eq.(\ref{eq:Pbar}) are then easily calculated with the use
of (\ref{eq:dP}):\ $\overline{\delta P^{n}}=(c_{1}^{n}+c_{2}^{n})/(n\lambda
l)$ for $n=2-5.$ Further simplification is possible if we assume that the
characteristic length $\sqrt{g}\gg d_{at}$, while{\bf \ }$\alpha \sim d_{at},
$ meaning that we have a small parameter $\alpha /\lambda g\ll 1$. In this
case the terms $K_{2,\ldots ,5}$ contain higher powers of the small
parameter $\alpha /\lambda g$ and can indeed be omitted. Expanding the
results in terms of the small parameter $\alpha /\lambda g\ll 1$, we find 
\begin{eqnarray}
\check{A} &=&A+a/l, \\
W &=&w_{1}(1-\alpha _{1}/\lambda g)+w_{2}(1-\alpha _{2}/\lambda g), \\
a &=&\alpha _{1}+\alpha _{2}+\beta , \\
\check{B} &=&B+\Delta B,
\end{eqnarray}
where $\beta =3B(w_{1}^{2}+w_{2}^{2})/\left( 2g^{2}\lambda ^{3}\right)
+5C(w_{1}^{4}+w_{2}^{4})/\left( 4g^{4}\lambda ^{5}\right) ,$ $\Delta
B=5C(w_{1}^{2}+w_{2}^{2})/(g^{2}\lambda ^{3}l).$ The terms $b$ and $f$ have
no effect on a behavior near the phase transition, since they are
proportional to $d_{at}/l$, and can be shown to be small. They can be
neglected together with the renormalization of the coefficient $C.$ The
dielectric function is 
\begin{equation}
\epsilon =1+\frac{4\pi }{A+a/l+3\check{B}\bar{P}^{2}+5C\bar{P}^{4}},
\end{equation}
with $\bar{P}$ determined from Eq.(\ref{eq:PbarGen}){\bf \ }and has a smooth
peak in all systems with nonsymmetric electrodes. In a symmetric case ($W=0)$
the second order transition is shifted down proportionally to $1/l.$ Indeed,
in the Landau theory $A=(T-T_{c0})/T_{0},$ where $T_{0}\sim T_{at}\sim 10^{5}
$K in displacive ferroelectrics, $T_{c0}$ the phase transition\ temperature
in the bulk, and the transition temperature in the film is 
\begin{equation}
T_{c}-T_{c0}=-aT_{0}/l.
\end{equation}
This result is similar to the one after Kretschmer and Binder \cite{Binder79}%
, but what is new here is that the shift depends on surface dipoles via the
new term\ $\beta $\ that pushes the transition temperature down. Indeed, $B$
is negative but small for weak first order transitions as in perovskites,
while $C$ has the usual ``atomic'' value, $C\sim P_{at}^{-4}$. The first
order phase transition is pushed closer to the second order, because of
positive renormalization of the coefficient $B\rightarrow \check{B}.$ The
shift of temperature of the first order transition in symmetric case is
defined mainly by the renormalization of the coefficient $A.$ Indeed, the
condition of the first order transition $\check{A}_{{\rm I}}=3\check{B}%
^{2}/16C$ can be presented as 
\begin{equation}
A_{{\rm I}}=A_{{\rm I}bulk}-a/l+3B\Delta B/8C,  \label{Renfirstor}
\end{equation}
where $A_{\text{{\rm I}}bulk}$ corresponds to the transition in the bulk and
we have neglected the term $\propto l^{-2}$. Let us emphasize that if,
according to Eq.(\ref{Renfirstor}), $A_{{\rm I}}$ becomes negative, this
equation may become inapplicable since the system can split into domains. It
has been shown in Ref.~\cite{Chensky82} that the ferroelectric transition in
a film with perfect metallic electrodes and a ``dead layer'' (e.g. vacuum
layer) proceeds with the domain formation even with dead layers down to
atomic thicknesses. The imperfect screening by metallic electrodes will
produce similar results. The strongly polarized surface regions that we
consider here are analogous to those dead layers. The value of $A$
corresponding to the transition with the domain formation depends on the
materials parameters and can be roughly estimated as $A\sim -d_{at}/\left(
l\epsilon ^{1/2}\right) $, where $\epsilon $ is the dielectric constant in
the direction perpendicular to the polar axis\cite{Chensky82,BLinh}. For
perovskites, which are uniaxial due to misfit strain, the value of $\epsilon 
$ can be large, so that the condition for the domain formation is $%
A<-d_{at}/\left( l\epsilon ^{1/2}\right) $ can be met in a tiny temperature
interval just below the loss of stability of the paraelectric phase at the
point $A=0.$ Therefore, in experiments with symmetric electrodes one would
see a phase transition very near $A=0,$ which is close to the transition
temperature in the bulk \cite{Saad04}. 

The present theory may, at least qualitatively, explain the observed
surprisingly strong smearing of the phase transitions in thin films, both
epitaxial and polycrystalline\cite{TFexp,Lookman04}. Indeed, we made the
estimate of the dielectric constants for Ba$_{1-x}$Sr$_{x}$TiO$_{3}$ ($x=0.3)
$ for thicknesses $l=1200$ and $775$ nm, which is in good semiquantitative
agreement with the recent data \cite{Lookman04}(Fig. 1). The value of the
surface bias field has been in the range $0.07-0.1$ CGSE, which corresponds
rather accurately to our above order-of-magnitude estimates. Slower fall-off
in the data for the dielectric constant at temperatures below the peak may
be related to usual presence of domains in this temperature region. It is
worth mentioning that there have been various hypotheses put forward to
explain the observed very strong smearing of the phase transition, like a
``dead'' layer or the relaxor behavior\cite{TFexp}. We would like to point
out that both of those explanations are not likely in the case of
epitaxially grown thin films. In particular, the relaxor model would require
a presence of strong nanometer scale disorder, whereas in e.g. epitaxial
PZT\ films on strontium ruthenate there is practically an ideal atomic
registry to the substrate\cite{Tybell01}. 

\begin{figure}[t]
\epsfxsize=3in 
\epsffile{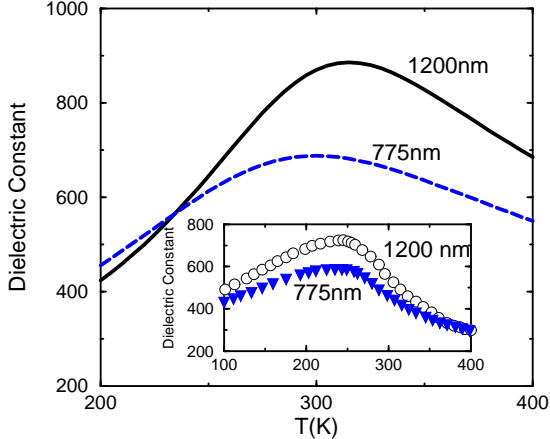 }
\caption{ Theoretical dielectric constants for films of Ba$_{0.5}$Sr$_{0.5}$%
TiO$_{3}$ with thicknesses $l=1200$ and 775nm showing complete smearing of
the phase transition as a result of the surface bias field (see text for
discussion). Inset: the data from Ref.~[21]. }
\label{fig:epsilon}
\end{figure}
The surface polarization discussed above is a special case of a polarization
due to gradients of a scalar quantity (concentration $c$ of e.g. oxygen
vacancies, density, temperature, etc.) and they are accounted for by a term
like 
\begin{equation}
f_{c}=-\gamma \vec{P}\vec{\nabla}c,  \label{eq:fgrad}
\end{equation}
in the free energy, where the coefficient $\gamma $\ is estimated as $\gamma
\sim P_{at}d_{at}$ \cite{Kogan64,LevMin80,Tagrev91}. This field should be
taken into account in the case of a film with a compositional profile
(grading)\ given by e.g. the concentration of one of the components of a
ferroelectric alloy $c=c(z)$. Generally, the other coefficients of the
thermodynamic potential also become inhomogeneous.{\bf \ }There are FE
systems with intentional concentration profile (graded)\cite{graded}\ and
unintentional (e.g. with a defect concentration profile) that are currently
a focus of research. The equation of state of the graded ferroelectric film
with $c=c(z)$ is 
\begin{eqnarray}
&&A(z)P+BP^{3}+CP^{5}-g\frac{d^{2}P}{dz^{2}}-D\nabla _{\perp }^{2}P 
\nonumber \\
&=&E_{0}+4\pi (\bar{P}-P)+\gamma \frac{dc}{dz}.  \label{eq:AzgamES}
\end{eqnarray}
If it were possible to neglect the inhomogeneity in other coefficients of
the thermodynamic potential, we have to simply add an average value of the
``field'' $\gamma dc/dz\sim \gamma \Delta c/l$ to the bias field $(\tilde{w}%
_{1}+\tilde{w}_{2})/l,$ where $\Delta c$ is the difference of concentration
through the sample ($\Delta c_{max}=1$). Account for the inhomogeneities of
the Landau coefficients makes the problem more difficult because even small
inhomogeneities of the coefficients have been shown to lead to the domain
formation if the bias field is absent \cite{BLinh}. As above, the main
effect of the surface bias fields is smearing out of the phase transition
into a monodomain state. However, a phase transition with formation of a
domain structure remains a possibility. For a special case of step-wise
concentration{\bf \ }we have found elsewhere \cite{BLcm04} that (i)\ in the
case of symmetric boundaries our previous results \cite{BLinh} are not
changed essentially by the presence of the interface field $\gamma $ and the
soft part splits into domains. In the case (ii) of asymmetric boundaries the
domain formation is possible but only for a much larger compositional
inhomogeneity. 

We thank J.M. Gregg, A. Kholkin, and J.F. Scott for stimulating discussions
and sharing their data.


\begin{references}
\bibitem{Landau37}  L.D. Landau, Phys. Zs. Sowjet. {\bf 11}, 26; {\bf 12, }%
123 (1937), in:\ {\it Collected Papers of L.D. Landau}, edited by D. ter
Haar (Gordon and Breach, New York 1965), pp.193, 233.

\bibitem{scott00}  J.F. Scott, {\it Ferroelectric Memories} (Springer, New
York, 2000).

\bibitem{nanofe}  J. J. Urban {\it et al}., Adv. Mater., 15, 423 (2003); Y.
Luo {\it et al.}, Appl. Phys. Lett. {\bf 83}, 440 (2003); I. Szafraniak {\it %
et al}.,{\it \ ibid.} {\bf 83}, 2211 (2003); M. Alexe and J.F. Scott, Key
Engineer. Mat. {\bf 206-213}, 1267 (2002).

\bibitem{GinzburgLandau}  V.L. Ginzburg and L.D. Landau, Zh. Eksp. Teor.
Fiz. {\bf 20}, 1064 (1950).

\bibitem{GPit58}  V.L. Ginzburg and L.P. Pitaevskii, Zh.\ Eksp. Teor. Fiz. 
{\bf 34}, 1240 (1958) [Sov. Phys. JETP, {\bf 7}, 858 (1958)].

\bibitem{degennes}  P.G. de Gennes, {\it Superconductivity of Metals and
Alloys} (Benjamin, New York, 1966).

\bibitem{Kaganov}  M.I. Kaganov and A.N. Omelyanchouk, Zh. Eksp. Teor. Fiz. 
{\bf 61}, 1679 (1971) [Sov.Phys. JETP {\bf 34, }895 (1972)].

\bibitem{Binder72}  K. Binder and P.C. Hohenberg, Phys. Rev. B {\bf 6, }3461
(1972).

\bibitem{Binder79}  R. Kretschmer and K. Binder, Phys. Rev. B {\bf 20}, 1065
(1979).

\bibitem{Tilley2001}  L.-H. Ong, J. Osman, and D.R. Tilley, Phys.\ Rev. B 
{\bf 63}, 144109 (2001).

\bibitem{LevSig88}  A.P. Levanyuk and A.S. Sigov, {\it Defects and
Structural Phase Transitions} (Gordon and Breach, New York, 1988).

\bibitem{bardeen36}  J. Frenkel, Z. f. Physik {\bf 51}, 232 (1928); J.
Bardeen, Phys. Rev. {\bf 49}, 653663 (1936).

\bibitem{Monch}  W. M\"{o}nch, {\it Semiconductor Surfaces and Interfaces}
(Springer, Berlin, 1995).

\bibitem{vbilt01}  B. Meyer and D. Vanderbilt, Phys. Rev. B {\bf 63}, 205426
(2001); R.E. Cohen, Ferroelectrics {\bf 194}, 323 (1997); Ph. Ghosez and
K.M. Rabe, Appl. Phys. Lett. {\bf 76}, 2767 (2000); J.B. Neaton and K.M.
Rabe, Appl. Phys. Lett. {\bf 82}, 1586 (2003).

\bibitem{Chensky82}  E.V. Chensky and V.V. Tarasenko, Sov. Phys. JETP {\bf 56%
}, 618 (1982) [Zh. Eksp. Teor. Fiz. {\bf 83}, 1089 (1982)].

\bibitem{BL}  A.M. Bratkovsky and A.P. Levanyuk, Phys. Rev. Lett. {\bf 84},
3177 (2000).

\bibitem{StemmerPTO95}  S. Stemmer {\it et al}., Phil. Mag. {\bf 71}, 713
(1995).

\bibitem{BLinh}  A.M. Bratkovsky and A.P. Levanyuk, Phys. Rev.\ B {\bf 66},
184109 (2002).

\bibitem{Saad04}  M. M. Saad {\it et al}., cond-mat/0406197.

\bibitem{TFexp}  Yu.A. Boikov and T. Claeson, J. Appl. Phys. {\bf 89}, 5053
(2001); Z. Kighelman {\it et al}., J. Appl. Phys. {\bf 91}, 1495 (2002);
C.B. Parker {\it et al.}, Appl. Phys. Lett. {\bf 81}, 340 (2002).

\bibitem{Lookman04}  A. Lookman {\it et al.}, J. Appl. Phys.{\bf 96}, 555
(2004).

\bibitem{Tybell01}  C.H. Ahn {\it et al.}, Science {\bf 276}, 1100 (1997);
P. Paruch {\it et al}. Appl. Phys. Lett. {\bf 79}, 530 (2001).

\bibitem{Kogan64}  Sh.M. Kogan, Fiz. Tverd.Tela (Leningrad) {\bf 5, }2829
(1963) [Sov. Phys. Solid State {\bf 5}, 2069 (1964)].

\bibitem{LevMin80}  A.P. Levanyuk and S.A. Minyukov, Fiz. Tverd. Tela
(Leningrad) {\bf 22, }1808 (1980).

\bibitem{Tagrev91}  A.K. Tagantsev, Phase Transitions {\bf 35}, 119 (1991).

\bibitem{graded}  N.W. Schubring {\it et al.}, Phys. Rev. Lett. {\bf 68},
1778 (1992); J.V. Mantese {\it et al.}, Appl. Phys. Lett. {\bf 71}. 2047
(1997); F. Jin {\it et al.}, Appl. Phys. Lett. {\bf 73}, 2838 (1998); M.S.
Mohammed {\it et al.}, J. Appl. Phys. {\bf 84}, 3322 (1998).

\bibitem{BLcm04}  A.M. Bratkovsky and A.P. Levanyuk, cond-mat/0402100.
\end{references}
\end{document}